\documentclass[12pt]{article}
\usepackage{amssymb,amsmath,epsfig}
\usepackage{caption}
\usepackage{subcaption}
\usepackage{epstopdf}
\graphicspath{ {D:/pics/} }
\allowdisplaybreaks
\begin{document}

\title{\bf The Impact of $f(\mathcal{G}, T)$ Gravity on the Evolution of Cavity in the Cluster of Stars}

\author{Rubab Manzoor $^1$, \thanks{rubab.manzoor@umt.edu.pk;~~dr.rubab.second@gmail.com}
M. Awais Sadiq $^1$ \thanks{awaismalhi007@gmail.com} and Imdad Hussain $^1$
\thanks{imdadhussainpu@gmail.com}\\
$^1$ Department of Mathematics,\\ University of Management and
Technology\\
Johar Town Campus,\\ Lahore, Pakistan.}
\date{}
\maketitle

\begin{abstract}
This paper analyzes the evolution of cavities for the cluster of
stars in the context of modified Gauss-Bonnet gravity. For this purpose, we
assume a spherically symmetric geometry with locally anisotropic
fluid distribution. It is assumed that the proper radial distance among
neighboring stellar components stays unchanged during purely areal
evolution stage. We provide some analytical solutions by using general
formulism in $f(\mathcal{G}, T)$ gravitation theory. The thick-shells
cavities at one or both boundary surfaces are found to satisfy the
Darmois conditions. Moreover, we also investigate the physical behavior
of cavity models by considering the stellar $4U 1820-30$. We conclude
that the dark matter has a strong impact on the evolution of cavities in the cluster of stars.
\end{abstract}

\section{INTRODUCTION}
The phenomenon of the cosmic expansion is a big breakthrough in this new age of
astrophysics. Results originating from cosmic microwave background radiation, BICEP
and Type Ia Supernova \cite{perl1}-\cite{koma1} have confirmed that there is an
accelerating expansion in our universe.
Dark energy is considered to be the main
cause of the expansion of the universe.
In accord with the recent results, originating
from Planck's estimations \cite{1}, it is discovered that approximately $95{\%}$ of
the cosmos is unknown and made up of $68{\%}$ of dark energy, $27{\%}$ of dark matter
(DM) and the remaining is normal matter. Dark matter is a hypothetical form of matter which
does not
or very weakly interact with electromagnetic radiation. It can be only observe
through its gravitational effects upon seen stellar. The phenomena of gravitational lensing,
rotational curves, mass discrepancy problem in galaxies show the existence and importance of DM
in the stellar evolution.

In this context, extended gravitational theories (EGT) emerge to unravel the mystery of our cosmos.
The EGT is formed
after generalizing the usual Einstein-Hilbert action to study the accelerated transformation
as well as the dark part of the universe. Nojiri and Odintsov \cite{2} clarified how this EGT are vital for studying
the transformation stages of the cosmos. The gravitational theories like $f(\mathbb{\phi})$, $f(R)$, $f(\mathcal{G})$,
$f(\mathcal{G},T)$ and $f(R,T)$ etc., were developed \cite{3}-\cite{15}
after being motivated by the standard Einstein-Hilbert Lagrangian.

One of the most discussed and recently developed theory of gravity is modified Gauss-Bonnet (MGB) gravity. It is believed that cosmological expansion can be detected in this theory of gravity because of the emergence of de Sitter point \cite{17}-\cite{21}. One of the most impressive nature of this gravity is that the existence of the Gauss-Bonnet invariant can prevent indistinct implications and unify the gravitational action \cite{22}. The MGB gravity \cite{15} was extended by combining the Gauss-Bonnet invariant with the trace of the stress-energy tensor. Shamir and Ahmad \cite{24} generated various cosmological possible models of MGB theory to identify the exact solutions by manipulating the Noether symmetry approach for the flat Friedmann-Robertson-Walker (FRW) universe. Sharif and Ikram \cite{25} measured the stability of certain linear perturbed models in the FRW universe and reviewed some relevant cosmological solutions in the field of MGB gravity. In another context, the same authors \cite{26} explored the static spherically symmetric traversable wormhole solutions in the framework of $f(\mathcal{G},T)$ gravity and deduced that the traversable wormhole solutions are physically acceptable in this theory. The exact solutions of the field equations in MGB theory for locally rotationally symmetric Bianchi type I anisotropic spacetime were evaluated by Shamir \cite{28}. Shamir and Sadiq \cite{sharhas} investigated the evolutionary phases of the universe via scalar functions in the field of MGB gravity. Shamir and Ahmad \cite{27} investigated the Noether symmetries for locally rotationally symmetric Bianchi type I and exerted the anisotropic effect to analyze $f(\mathcal{G}, T)$ gravity cosmological models. Sharif and Yousaf \cite{29} worked on the modification of stellar bodies in MGB gravity with some definite models and deduced that the star compactness increases for various models of gravity.

A cluster of stars is a collection of stars that are gravitationally linked with each other. The average gap between stars in a cluster is significantly less than the average gap between remaining galaxies. The study of star clusters is thought to be useful in examining the formation and evolution of galaxies. There have been noteworthy theoretical attempts to investigate evolution as well as the formation of star clusters \cite{30}. Ipser and 
Thorne \cite{isper} investigated the relativistic, spherically symmetric star clusters and found that the relativistic instability strongly influences the role of star clusters. Kruijssen \textit{et al.} \cite{krui} discussed the formation as well as the evolution of star clusters and deduced that the cluster of stars is the tracers of galaxy evolution. Krumholz \textit{et al. }\cite{krum} studied the cluster of stars across cosmic time and provided a life cycle of star clusters with several transformation stages. Recently, Manzoor and Shahid \cite{rubab} examined the dynamics of evolving star clusters in $f(R)$ gravity with the help of scalar functions. The DM is a fundamental component of the cluster of stars and  it is believed
to be about $95 {\%}$ of the total mass of the stars cluster.

Cosmic voids are huge gaps between galactic filaments that are very sparsely populated. Voids have typical sizes of hundreds of millions of light-years and have occupied about $90{\%}$ of known space. Voids are not entirely empty but due to their low density, the stripping of galactic gas is assumed to be insufficient in galaxy groups. In this way, voids provide a different scenario to study the formation and evolution of the stellar system. It is assumed that the visible large-scale structure of the universe is sponge-like, controlled by voids \cite{Wilt}. Hoyle and Vogeley \cite{hoyle} presented an analysis of voids in the 2dF Galaxy Redshift Survey and identified the void regions as well as the measurement of void statistics. Tikhonov and Karachentsev \cite{tikh} discovered minivoids and found their expansion volume. Moreover, supervoids of different scales have also been found \cite{rud}. Many scientists have also discovered that galaxies with void centers have a poor tendency to have greater rates of star formation. It is vital to remember that voids are not always empty or spherical. In simulation results or deep redshift studies, voids are typically shown as vacuum spherical cavities neighboured by a fluid.

On the other hand, the cavity is a hollow depression in the structure of an astronomical object and can be assumed as a void
precursor. Skripkin \cite{31} was the first to raise the topic of cavity evolution followed by a central explosion in a spherically symmetric distribution. Subsequently, by using Skripkin conditions, a Minkowskian cavity arises around the center of the fluid distribution. This problem has been thoroughly investigated, revealing that under Skripkin conditions, the scalar expansion vanishes for isotropic fluid with constant energy density distribution \cite{32}. Further investigation into this topic revealed the requirement of cavity existence inside the system in order to get the vanishing expansion scalar. Herrera \textit{et al.} \cite{33} investigated that the Skripkin model is inconsistent with the Darmois junction conditions \cite{34} and showed that in general, the expansion-free condition for any non-dissipative fluid requires the inhomogeneous distribution of energy density. Herrera \textit{et al.} \cite{35} studied the evolution of cavity within spherically symmetric relativistic fluids by assuming that proper radial distance among adjacent fluid particles remains constant during their evolution. Sharif and Bhatti \cite{36} investigated the shear-free and cavity models with plane symmetry. Moreover, Yousaf and Bhatti \cite{37} studied the cavity evolution and instability constraints of relativistic interiors and found that the dark sources due to higher-order gravity model is responsible for the instability of the system.

In this paper, we will discuss the evolution of the cavity in $f(\mathcal{G}, T)$ gravity for a spherically symmetric cluster of stars. We apply the method of purely areal evolution wherein the change in the proper radial distance among any two infinitesimally close particles of fluid per unit of proper time disappears \cite{35}. We are mainly interested in the evolution of the cavity after it has been created, not in the conditions and dynamics that caused it to arise. In this context, we suppose that the two hypersurfaces are constraining the fluid. The inner hypersurface separates the cavity within which Minkowski spacetime exists and the outer one splits the fluid corrections from Vadiya or Schwarzchild spacetime (based on whether we assume dissipative or adiabatic evolution). In section \textbf{2}, we provide the formulation of modified MGB gravity field equations. Section \textbf{3} discusses the spherically symmetric cluster of stars in $f(\mathcal{G},T)$ theory. Section \textbf{4} discusses briefly the boundary conditions on the inner as well as outer boundary surfaces. Section \textbf{5} provides the physical interpretation of the purely areal and radial evolution conditions. In section \textbf{6}, different models of cavities are presented. Section $\textbf{7}$ is devoted to discussing some cases of cavities satisfying Darmois conditions on both hypersurfaces. In section $\textbf{8}$, we provide the physical interpretation of cavity models in the context of the MGB gravity model. The concluding remarks are discussed in the last section.

\section{MGB Fields Equations}

The MGB gravity is given by the action \cite{15}
\begin{equation}\label{f}
\mathcal{A}_{f(\mathcal{G},T)}=\frac{1}{2\kappa^2}
\int[f(\mathcal{G},T)+R]\sqrt{-g}d^4x+ \int\sqrt{-g}\mathfrak{L}_M d^4x,
\end{equation}
where $\mathcal{G}$ indicates the Gauss-Bonnet invariant, $T$ shows stress energy tensor, $g$ is the
determinant of metric tensor, $R$ denotes the Ricci scalar, $\mathfrak{L}_M$ shows the standard Lagrangian
coupled with matter and $\kappa$ signifies the coupling constant. Varying the action (\ref{f}) according to metric tensor \cite{15}
\begin{equation}\label{4b}
\begin{split}
R_{\lambda\mu}-\frac{1}{2}Rg_{\lambda\mu} &= \kappa^2 T_{\lambda\mu}
-(T_{\lambda\mu}+
\Theta_{\lambda\mu})f_{T}(\mathcal{G},T)+\frac{1}{2}g_{\lambda\mu}
f(\mathcal{G},T)-(2RR_{\lambda\mu}\\ &-4R^\xi_\lambda
R_{\xi\mu}-4R_{\lambda\xi\mu\eta}R^{\xi\eta}+2R^{\xi\eta\delta}_\lambda
 R_{\mu\xi\eta\delta})f_{\mathcal{G}}(\mathcal{G},T)-(2Rg_{\lambda\mu}\nabla^2\\ &+R^\xi_\lambda\nabla_\mu\nabla_\xi
-4R_{\lambda\mu}\nabla^2-4g_{\lambda\mu}R^{\xi\eta}\nabla_\xi\nabla_\eta-2R\nabla_\lambda\nabla_\mu
\\ &+4R_{\lambda\xi\mu\eta}\nabla^\xi\nabla^\eta+4R^\xi_\mu\nabla_\lambda\nabla_\xi
)f_{\mathcal{G}}(\mathcal{G},T),
\end{split}
\end{equation}
where the subscript $\mathcal{G}$ or $T$ denote the partial derivatives and $\nabla_\lambda\nabla^\lambda=\Box=\nabla^2$ represents the
d'Alembert operator. It is interesting to notice that usual general relativity equations are
recovered by substituting $f(\mathcal{G},T)=0$ in Eq.(\ref{4b}). In
addition to that, the field equation of $f(\mathcal{G},T)$ gravity are simplified
to the field equation of $f(\mathcal{G})$ gravity by replacing
$f(\mathcal{G},T)$ with $f(\mathcal{G})$. The trace of
Eq.(\ref{4b}) is
\begin{eqnarray}\nonumber
\begin{split}
&T+R+2\mathcal{G}f_{\mathcal{G}}(\mathcal{G},T)-(\Theta+T)f_{T}(\mathcal{G},T)-2R\nabla
^2f_{\mathcal{G}}(\mathcal{G},T)\\ &+4R^{\lambda\mu}\nabla_\lambda\nabla_\mu
f_{\mathcal{G}}(\mathcal{G},T)=0.
\end{split}
\end{eqnarray}
The covariant divergence of Eq.(\ref{4b}) is given by
\begin{equation}\nonumber
\begin{split}
\nabla^\lambda
T_{\lambda\mu}=\frac{f_{T}(\mathcal{G},T)}{1-f_{T}(\mathcal{G},T)}\Bigg[(T_{\lambda\mu}
+\Theta_{\lambda\mu})\nabla^\lambda(\ln
f_{T}(\mathcal{G},T))+\nabla^\lambda
\Theta_{\lambda\mu}-\frac{1}{2}g_{\lambda\mu}\nabla^\lambda
 T\Bigg],
\end{split}
\end{equation}
where $\Theta_{\lambda\mu}$ is given as
\begin{equation}\label{5b}
\Theta_{\lambda\mu}=g^{\xi\eta}\frac{\delta T_{\xi\eta}}{\delta
g_{\lambda\mu}}.
\end{equation}
The term $\Theta_{\lambda\mu}$ can be obtained with the help of the following expression
\begin{equation}\label{6b}
\frac{\delta T_{\lambda\mu}}{\delta g^{\xi\eta}}=\frac{\delta
g_{\lambda\mu}}{\delta
g^{\xi\eta}}\mathfrak{L}_M+g_{\lambda\mu}\frac{\partial
\mathfrak{L}_M}{\partial g^{\xi\eta}}-2\frac{\partial^2
\mathfrak{L}_M}{\partial g^{\xi\eta}
\partial g^{\lambda\mu}}.
\end{equation}
Substituting Eq.(\ref{6b}) in Eq.(\ref{5b}), we get
\begin{equation}\label{7b}
\Theta_{\lambda\mu}=-T_{\lambda\mu}+g_{\lambda\mu}\mathfrak{L}_M
-2g^{\xi\eta}\frac{\partial^2 \mathfrak{L}_M}{\partial
g^{\lambda\mu}\partial g^{\xi\eta}}.
\end{equation}
The matter Lagrangian can be considered as
$\mathfrak{L}_{M}=-\rho$, so Eq.(\ref{7b}) yields the
 following form
\begin{equation}\label{9b}
\Theta_{\lambda\mu}=-2T_{\lambda\mu}-\rho g_{\lambda\mu}.
\end{equation}
The Einstein field equations (\ref{4b}) can be described in an identical form as
\begin{equation}\nonumber
G_{\lambda\mu}=\kappa^2 T_{\lambda\mu}^{eff},
\end{equation}
where
\begin{equation}\label{11b}
\begin{split}
T_{\lambda\mu}^{eff}&=(1+f_{T})T_{\lambda\mu}+\rho
g_{\lambda\mu}f_{T}+\frac{1}{2}g_{\lambda\mu}f(\mathcal{G},T)-(2RR_{\lambda\mu}
\\ &-4R^{\xi}_{\lambda}R_{\xi\mu}-4R_{\lambda\xi\mu\eta}R^{\xi\eta}
+2R^{\xi\eta\delta}_{\lambda}R_{\mu\xi\eta\delta})f_{\mathcal{G}}(\mathcal{G},T)
\\ &-(2Rg_{\lambda\mu}\nabla^2-2R\nabla_\lambda\nabla_\mu
-4g_{\lambda\mu}R^{\xi\eta}\nabla_\xi\nabla_\eta
-4R_{\lambda\mu}\nabla^2\\ &+4R^{\xi}_{\lambda}\nabla_\mu\nabla_\xi
+4R^{\xi}_{\mu}\nabla_{\lambda}\nabla_{\xi}+4R_{\lambda\xi\mu\eta}\nabla^{\xi}\nabla^{\eta})f_{\mathcal{G}}(\mathcal{G},T).
\end{split}
\end{equation}

The extra curvature configurations of MGB gravity can be instigated by considering
disconnected expressions for functions of $\mathcal{G}$ and $T$. In this study, we assume the following power law model
\begin{equation}\label{6969}
f(\mathcal{G}, T)=f_1 (\mathcal{G})+f_2 (T).
\end{equation}
  Here $f_1 (\mathcal{G})=\alpha \mathcal{G}^n$ and $f_2 (T)=\beta T$, with $\alpha$ and $\beta$ are arbitrary real numbers and assumed as dark source parameters. Cognola \textit{et al.} \cite{cogn1} suggested this model and it is fascinating because there are no chances of appearing finite singularities.  Furthermore, these models can be used as tools to understand the dark mysterious part of the universe. For ease of reference, we consider $n=2$ for our upcoming analysis.

\section{Fluid Distributions and Kinematical Variables}
We assume a non-static spherical cluster of stars composed of  matter and DM, where stars are supposed as fluid particles. The spherically symmetric distribution is supposed to be anisotropic, dissipative and is bounded by a spherical surface $\Sigma^{(e)}$. If we consider comoving coordinates inside $\Sigma^{(e)}$, then the interior geometry of star cluster  metric can be written as
\begin{equation}\label{12b}
ds^2_-=-X^2(t, r)dt^2+Y^2(t, r)dr^2+C^2(t, r)\left(d\theta^2+\sin^2\theta d\phi^2\right),
\end{equation}
where $X$ and $Y$ are assumed to be dimensionless, however $C$ has the same dimension as $r$. Equation (\ref{12b}) indicates that the proper radius inside $\Sigma^{(e)}$ spherical surface is given by $\int B dr$ and $C$ gives the areal radius. The stress energy tensor $T^{-{eff}}_{\lambda\mu}$ inside $\Sigma^{(e)}$ has the following form
\begin{equation}\label{13b}
\begin{split}
T^{-{eff}}_{\lambda\mu}=&(P^{eff}_\bot+\rho^{eff})V_\lambda
V_\mu+P^{eff}_\bot
g_{\lambda\mu}+(P^{eff}_r-P^{eff}_\bot)\chi_\lambda\chi_\mu
\\&+q^{eff}_\lambda V_\mu +V_\lambda q^{eff}_\mu,
\end{split}
\end{equation}
where  $P^{eff}_r$ and $P^{eff}_\bot$ signify the radial and
tangential pressure respectively, $\rho^{eff}$ represents the energy
density, $q^{\lambda({eff})}$ shows the dissipation in the form of heat flux,
four-unit-vector in the radial direction is denoted by $\chi^\lambda$ and
$V^\lambda$ is the four velocity of the fluid. These
quantities fulfill the following relations
\begin{equation}\label{14b}
V^\lambda V_\lambda=-1, \quad V^\lambda q^{eff}_\lambda=0, \quad
\chi^\lambda \chi_\lambda=1, \quad \chi^\lambda V_\lambda=0.
\end{equation}
Moreover, we can split the effective quantities in terms of matter and DM components as
\begin{eqnarray}\nonumber
\rho^{eff}&=&\rho^{M}+\rho^{D},
\\\nonumber
P^{eff}_{r}&=&P^{M}_{r}+P^{D}_{r},
\\\nonumber
P^{eff}_{\bot}&=&P^{M}_{\bot}+P^{D}_{\bot},
\\\nonumber
q^{eff}&=&q^{M}+q^{D}.
\end{eqnarray}
where DM and baryonic matter contributions are denoted by superscripts
$M$ and $D$ respectively.

The fluid expansion $\Theta$ and four-acceleration $a_\lambda$ are
\begin{equation}\label{15b}
\quad \Theta=V^\lambda_{;\lambda},~~
a_\lambda=V_{({\lambda;\mu})}V^\mu
\end{equation}
and the shear tensor $\sigma_{\lambda\mu}$ is
\begin{equation}\label{16b}
\sigma_{\lambda\mu}=V_{(\lambda;\mu)}+a_{(\lambda}V_
{\mu)}-\frac{1}{3}\Theta h_{\lambda\mu},
\end{equation}
where
\begin{equation}\label{17b}
h_{\lambda\mu}=V_{\lambda} V_{\mu}+g_{\lambda\mu}.
\end{equation}
Since we considered the metric (\ref{12b}) in comoving coordinates, so
\begin{equation}\label{18b}
V^\lambda=X^{-1}\delta^\lambda_0,
~~\chi^\lambda=Y^{-1}\delta^\lambda_1,~~ q^{\lambda(eff)}=q^{eff} Y^{-1}\delta^a_1,
\end{equation}
where $q^{eff}$ is function of $t$ and $r$. Equations (\ref{15b}) and
(\ref{18b}) are used to determine the non-zero component of four
acceleration as well as its scalar.
\begin{equation}\label{19b}
a_1=\frac{X^\prime}{X},  \quad  a=( a_\lambda a^\lambda)^\frac{1}{2}
=\Bigg(\frac{X^\prime}{XY}\Bigg),
\end{equation}
and the expansion scalar is given as
\begin{equation}\label{20b}
\Theta=\frac{1}{X}\Bigg(\frac{\dot{Y}}{Y}+2\frac{\dot{C}}{C}\Bigg).
\end{equation}
Here overdot and prime represent derivative according to $t$ and $r$, respectively. We get non-zero
components of the shear from Eqs.(\ref{16b}) and (\ref{18b}) as
\begin{equation}\label{21b}
\sigma_{11}=\frac{2}{3}Y^2\sigma, \quad\sigma_{22}=
\frac{\sigma_{33}}{\sin^2\theta}=-\frac{1}{3}C^2\sigma,
\end{equation}
and its scalar can be found as
\begin{equation}\label{22b}
\sigma_{\lambda\mu} \sigma^{\lambda\mu}=\frac{2}{3}\sigma^2,
\end{equation}
where
\begin{equation}\label{23b}
\sigma=\frac{1}{X}\Bigg(\frac{\dot{Y}}{Y}-\frac{\dot{C}}{C}\Bigg).
\end{equation}
The shear tensor can also be written in terms of projection tensor as
\begin{equation}\label{24b}
 \sigma_{{\lambda}{\mu}}=(\chi_\lambda \chi_\mu-\frac{1}{3}h_{\lambda\mu})\sigma.
\end{equation}
The line element (\ref{12b}) for model (\ref{6969}) has the following field equations
\begin{align}
&
\begin{aligned}[t]
8\pi \rho^{eff}=G_{00}=8\pi X^2\Bigg[\rho^M
+\frac{\alpha \mathcal{G}^2}{2}-\frac{\beta T}{2}-
\frac{\hat{\psi}_{00}}{X^2}\Bigg], \label{25b}
\end{aligned}
\\[\jot]
 &
\begin{aligned}[t]
8\pi q^{eff}=G_{01}=8\pi XY\Bigg[-(1+\beta)q^{M}-\frac{\hat{\psi}_{01}}{XY}\Bigg], \label{26b}
\end{aligned}
\\[\jot]
 &
 \begin{aligned}[t]
8\pi P_r^{eff}=G_{11}=8\pi Y^2\Bigg[\rho^M \beta+(1+\beta)P^M_r
-\frac{\alpha \mathcal{G}^2}{2}+\frac{\beta T}{2}-
\frac{\hat{\psi}_{11}}{Y^2}\Bigg], \label{27b}
\end{aligned}
\\[\jot]
 &
 \begin{aligned}[t]
8\pi P_\bot^{eff}=G_{22}=8\pi C^2\Bigg[\rho^M \beta+(1+\beta)P^M_\bot
-\frac{\alpha \mathcal{G}^2}{2}+\frac{\beta T}{2}-
\frac{\hat{\psi}_{22}}{C^2}\Bigg], \label{28b}
\end{aligned}
\end{align}
where $ \psi_{00},\psi_{01},\psi_{11}$ and $\psi_{22}$ are the dark components and the hat represents that these components are evaluated after using the MGB model (\ref{6969}). The expressions of $ \psi_{\alpha\beta}$ are given in the appendix. The stress energy tensor (\ref{13b}) can be rewritten in an alternative form as \cite{42}-\cite{43}
\begin{equation}
T^{-(eff)}_{\lambda\mu}=\rho^{eff}V_\lambda
V_\mu+\bar{P}^{eff}h_{\lambda\mu}+\Pi^{eff}_{\lambda\mu}+q^{eff}(V_\lambda
\chi_\mu+\chi_\lambda V_\mu),
\end{equation}
 with
\begin{eqnarray}\nonumber
\bar{P}^{eff}=\frac{1}{3}h_{\lambda\mu}T^{\lambda\mu({eff})}=\frac{P^{eff}_r+2P^{eff}_\bot}{3},
\end{eqnarray}
\begin{eqnarray}\nonumber
\Pi^{{\lambda\mu}({eff})}=(h^{(\lambda}_{\gamma}
h^{\mu)}_\delta-\frac{1}{3}h^{\lambda\mu}
h_{\gamma\delta})T^{{\gamma\delta}({eff})}=\Pi^{eff}(\chi^{\lambda}
\chi^{\mu}-\frac{1}{3}h^{\lambda\mu}),
\end{eqnarray}
and
\begin{eqnarray}\nonumber
\Pi^{eff}=P^{eff}_r-P^{eff}_\bot.
\end{eqnarray}
We now consider the mass function $m(t,r)$ introduced by Misner and Sharp \cite{44} (also see \cite{45}), which reads
\begin{equation}\label{29b}
m=\frac{R^3}{2}{R_{23}}^{23}=\frac{C}{2}\Bigg[\Bigg(\frac{\dot{C}}{X}\Bigg)^2
+1-\Bigg(\frac{C^\prime}{Y}\Bigg)^2\Bigg].
\end{equation}
The variations of areal radius with respect
to proper time $T$ may be used to explain the velocity $U$ of a
collapsing fluid.
\begin{equation}\label{31b}
U=D_TC<0,
\end{equation}
where $D_T=\frac{1}{X}\frac{\partial}{\partial t}$. Equations (\ref{29b}) and (\ref{31b}) provide
\begin{equation}\label{32b}
E=\frac{C^\prime}{Y}=\Bigg(U^2-\frac{2m}{C}+1\Bigg)^{\frac{1}{2}}.
\end{equation}
Using field equations with (\ref{31b}) and (\ref{32b}) (details in \cite{32}), we derive the following equation from Eq.(\ref{29b})
\begin{equation}\label{33b}
m^\prime=4\pi \Bigg(\rho^{eff}+q^{eff}\frac{U}{E}\Bigg)C^{\prime}C^2,
\end{equation}
which implies that
\begin{equation}\label{34b}
m=4\pi \int^r_0\Bigg(\rho^{eff}+q^{eff}\frac{U}{E}\Bigg)C^2C^{\prime}dr.
\end{equation}
We assumed that the distribution has a normal center, thus,
$m(0)=0$. The Weyl tensor is useful to describe the tidal forces effects upon the stellar.
The terms $E_{\lambda\mu}$ and $H_{\lambda\mu}$ represent the electric as well as magnetic parts
of the Weyl curvature tensor. The
magnetic component of the Weyl tensor disappears
in the case of spherical symmetry and the electric part is defined as
\begin{equation}\label{35b}
E_{\lambda\mu}=C_{\lambda\rho\mu\gamma}V^\rho V^\gamma ,
\end{equation}
which can be written as
\begin{equation}\label{36b}
E_{\lambda\mu}=\varepsilon(\chi_\lambda \chi_\mu-
\frac{1}{3}h_{\lambda\mu}),
\end{equation}
where
\begin{equation}\label{37b}
\begin{split}
\varepsilon&=\frac{1}{2X^2}\Bigg[\frac{\ddot{C}}{C}-
\frac{\ddot{Y}}{Y}-\Bigg(\frac{\dot{C}}{C}-\frac{\dot{Y}}{Y}\Bigg)
\Bigg(\frac{\dot{X}}{X}+\frac{\dot{C}}{C}\Bigg)\Bigg]-\frac{1}{2C^2}\\ &+\frac{1}{2Y^2}
\Bigg[\frac{X''}{X}-\frac{C''}{C}
+\Bigg(\frac{Y'}{Y}+\frac{C'}{C}\Bigg)\Bigg(\frac{C'}{C}-\frac{X'}{X}\Bigg)\Bigg].
\end{split}
\end{equation}
We can rewrite $\varepsilon$ by using the mass function (\ref{29b})
and field equations as (see \cite{32} for details)
\begin{equation}\label{38b}
\begin{split}
\varepsilon=4\pi(\rho^{eff}-P_r^{eff}+P_\bot^{eff})-\frac{3m}{C^3}.
\end{split}
\end{equation}

\section{The Exterior Geometry and Boundary Conditions}
We consider the Vaidya spacetime
(or Schwarzschild in the non-dissipative case) outside $\Sigma^{(e)}$, which states that
all emitted radiation has no mass, defined as
\begin{equation}\label{39b}
ds^2=-\Bigg[1-\frac{2M(\upsilon)}{r}\Bigg]d\upsilon^2-2drd\upsilon+r^2(d\theta^2+
sin^2\theta d\phi^2).
\end{equation}
Here, $\upsilon$ describes the retarded
time and $M(\upsilon)$ shows the total mass.
Darmois junction conditions are used in order to
form a match between the non-adiabatic sphere and Vaidya's
spacetime on the surface $r$=constant=$r_{\Sigma^{(e)}}$,
in absence of thin shells \cite{41}, \cite{46}-\cite{48}.
These junction conditions
incorporates the uniformity of first and second essential
forms based on matching hypersurfaces. We get the
following results from Darmois conditions
\begin{equation}\label{40b}
m(t,r)^{\Sigma^{(e)}}_=M(\upsilon),
\end{equation}
\begin{equation}\label{41b}
\begin{split}
&2\Bigg(\frac{\dot C'}{C}-\frac{X'}{X}\frac{\dot C}{C}-\frac{\dot
C}{C}\frac{\dot
Y}{Y}\Bigg)=\frac{X}{Y}\Bigg[\frac{C^\prime}{C}\Bigg(2\frac{X^\prime}{X}
+
\frac{\dot{C}}{C}\Bigg)-\Bigg(\frac{Y}{C}\Bigg)^2\Bigg]-\\ &\frac{Y}{X}\Bigg[2\frac{\ddot{C}}{C}-\frac{\dot{C}}{C}\Bigg(2\frac{\dot{X}}{X}
-\frac{\dot{C}}{C}\Bigg)\Bigg],
\end{split}
\end{equation}
and
\begin{equation}\label{42b}
q^{eff} { ^{{\Sigma^{(e)}}}_{=}}\frac{L}{4{\pi}r}.
\end{equation}
Here ${ ^{{\Sigma^{(e)}}}_{=}}$ signifies that both sides of the
equation are estimated on ${\Sigma^{(e)}}$ and $L$ is the total
luminosity of the cluster, given as
\begin{equation}\label{43b}
L=L_\infty\bigg(2\frac{dr}{d\upsilon}-\frac{2m}{r}+1\bigg)^{-1},
\end{equation}
with
\begin{equation}\label{44b}
L_\infty=\frac{dM}{d\upsilon},
\end{equation}
is the total luminosity at
infinity and evaluated by a stationary observer in the state of rest. Equation (\ref{41b}) and field
equations provide
\begin{equation}\label{45b}
P^{eff}_r{^{{\Sigma^{(e)}}}_{=}}q^{eff}.
\end{equation}
In the cavity formation case, to delimit the cavity on the boundary
surface, there must be matching conditions among the solution and Minkowski
spacetime on the boundary surface. If we refer $\Sigma^{(i)}$ as the boundary surface between the cluster fluid and cavity, then the matching corrections of Minkowski
spacetime provide
\begin{equation}\label{46b}
m(t,r)^{ \Sigma^{(i)}}_=0,
\end{equation}
\begin{equation}\label{47b}
P^{eff}_r{^{{\Sigma^{(i)}}}_{=}}q^{eff}.
\end{equation}
Moreover, if we suppose our cavity to be empty,
then we have $L{^{\Sigma^{(i)}}_{=}} 0$, which gives
\begin{equation}\label{48b}
P^{eff}_{r}{^{{\Sigma^{(i)}}}_{=}}q^{eff}
{^{{\Sigma^{(i)}}}_{=}}0.
\end{equation}
It should be noticed here, that for the existence of a thin shell on $\Sigma^{(e)}$ or
$\Sigma^{(i)}$, above conditions need to be rested and enable
the divergence within the mass function \cite{49}.

\section{The Purely Areal Evolution Condition and Radial Velocity}
Here we will discuss a new definition of collapsing velocity $U$. Previously, it was defined as, the change in the areal radius $C$
per unit proper time. The collapsing velocity $U$ can also be defined as, the change in the infinitely small proper radial distance between two adjacent points $(\delta l)$ per unit of proper time, that is, $D_T(\delta l)$. It is given as (see \cite{32} for details)
\begin{equation}\label{49b}
\frac{D_T (\delta l)}{\delta l}=\frac{1}{3}(\Theta+2\sigma).
\end{equation}
It can be seen from the above equation that this infinitesimal rate of change is related to expansion and shear effects.
Using Eqs.(\ref{20b}) and (\ref{23b}), we get
\begin{equation}\label{50b}
\frac{D_T (\delta l)}{\delta l}=\frac{\dot{Y}}{X{Y}}.
\end{equation}
From  Eqs.(\ref{20b}), (\ref{23b}), (\ref{31b}) and (\ref{50b}), we obtain
\begin{equation}\label{51b}
\sigma=-\frac{D_TC}{C}+\frac{D_T(\delta l)}{\delta
l}=-\frac{U}{C}+\frac{D_T(\delta l)}{\delta l},
\end{equation}
and
\begin{equation}\label{52b}
\Theta=\frac{2D_TC}{C}+\frac{D_T(\delta l)}{\delta
l}=\frac{2U}{C}+\frac{D_T(\delta l)}{\delta l}.
\end{equation}
Therefore, the areal velocity $U$ has an affiliation with the change of areal radius $R$ of a layer of fluid particle. On the other hand,
$D_T (\delta l)$, which can also be referred as velocity, is the relative velocity among the adjacent layers of fluid particles within the cluster. Equation (\ref{51b}) implies that, for $U<0$, the collapsing cluster will become shearless if $D_T(\delta l)<0$, which means the relative distance among the fluid particles diminish in such a way that it nullifies the values of $U$. According to Eq.(\ref{52b}), we can have the expansion-free case if areal velocity $U$ countervails the relative velocity $D_T(\delta l)$. Thus, in order to have a collapsing expansion-free situation, we should have
$U<0$ and $D_T(\delta l)>0$. In the case of outwardly flow ($U>0$), expansion-free is attainable if $D_T(\delta l)<0$. Furthermore, it has already been shown in \cite{35} that purely evolution condition is required for the existence of cavity ($D_T(\delta l)=0$ with $U\neq0$). Hence, in order to derive the cavity solutions in cluster, we consider the condition $D_T(\delta l)=0$ with $U\neq0$. This condition along with Eqs.(\ref{51b}) and (\ref{52b}) provides, $\Theta=-2\sigma$. Using the obtained result in Eq.(\ref{26b}), we get
\begin{equation}\label{53b}
\sigma^\prime+\frac{\sigma C^\prime}{C}=-\frac{4\pi q^{eff}
C^\prime}{E}.
\end{equation}
Integrating the above equation with respect to $r$
\begin{equation}\label{54b}
\sigma=\frac{\zeta(t)}{C}-\frac{4\pi}{C}\int^r_0
q^{eff}\frac{CC^\prime}{E}dr,
\end{equation}
where $\zeta(t)$ is a function of integration. It can be observed that, we should use the regularity
condition $\zeta = 0$, if all the spherical surface of star cluster including center $(r = 0)$ is contained with
the baryonic and non-baryonic fluid particles. Anyhow, we are interested in the formation of cavity neighboring the center, so such condition is not preferred. On the contrary, Eq.(\ref{54b}) with Eq.(\ref{51b}) provides
\begin{equation}\label{55b}
U=-\zeta+4\pi\int^r_0 q^{eff}\frac{CC^\prime}{E}dr.
\end{equation}
Therefore, in case of non-dissipation $(q^{eff}=0\Longrightarrow q^M=0=q^D)$, the purely areal evolution condition provides $U=U(t)$. This outcome is undoubtedly not suitable with a regular symmetry center except if $U=0$. However, if we need the purely areal evolution condition to be suitable with the time-dependent stage $(U\neq0)$, we suppose that either

\textbf{\textbf{.}} There is no symmetry center in the fluid.

or

\textbf{\textbf{.}} The center is surrounded by a compact spherical region of other spacetime,
perfectly matched to the rest of the fluid inside the cluster.

We will rule out the first option because we
are more intrigued in representing the local objects without unusual
topological structure of a cluster with no center. In
addition, we have considered an internal vacuum Minkowski spherical
vacuole in the context of the second option. It can be noticed that the dissipation due to DM ($q^D$) might be splurging of non-baryonic particles. This kind of dissipation occurs because of the gravitational energy of DM and
it cannot be ignored because of DM's gravitational effects.
Therefore, during the evolution of star cluster with DM, the non-dissipative case is very unique.

Now we assume another case, that is absolutely areal dissipative evolution
$(D_T(\delta l)=0)$. If there is no cavity surrounding the center and the gravitating fluid fills the entire spherical cluster, we get a symmetry center. In this case, we need to substitute $\zeta=0$ and then Eq.(\ref{55b}) gives
\begin{equation}\label{56b}
U=4\pi \int^r_0 q^{eff}\frac{CC^\prime}{E}dr,
\end{equation}
which is consistent with a regular symmetry center. In this
scenario, we will assume that the center is bounded by cavity in an \textit{ad hoc} manner. However, the following qualitative argument
hints at this assumption in some way.

All terms inside the integral are positive in case of an outgoing dissipation $(q^{eff}>0)$ and we can deduce from Eqs.(\ref{52b}) as well as (\ref{56b}) that $\Theta>0$ and $U>0$. Conversely, for inwardly dissipative condition ($q^{eff}<0$), we will get $\Theta<0$ and $U<0$. Since we are considering dissipation due to matter and DM, so one of the source of this dissipation is gravitational energy. Furthermore, according to Kelvin-Helmholtz phase of evolution \cite{51}, when outwardly dissipation phase occurs from gravitational energy then we should expect a contraction instead of an expansion. The above discussion indicates that the dissipation due to heat flux is different from dissipation because of gravitational energy. Hence, we can say that the outgoing dissipation due to DM ($q^D>0$) is the contraction phase of evolution, whereas, the inwardly dissipation ($q^D<0$) contributes to the expansion phase of evolving cluster. Thus, we deduce that purely areal evolution condition
are perfectly suited for describing the evolution of
a star cluster with cavity around the center.

Another way of representing the purely areal evolution condition in covariant form can be achieved by using Eqs.(\ref{51b}) and (\ref{52b}) in
Eq.(\ref{24b}) as
\begin{equation}\label{57b}
\sigma_{\lambda\mu}=\frac{\Theta}{2}(\chi_\lambda \chi_\mu-
\frac{1}{3}h_{\lambda\mu}).
\end{equation}

\section{Some Cavity Models}
In this section, we will explore some general characteristics of
models which satisfy the purely areal evolution condition. This idea is similar to the concept recommended by
Skripkin \cite{31}, that is, the explosion in the center causes an
overall expansion throughout the fluid, resulting in formation of a cavity around the center. But in this study, we are considering $D_T(\delta l)=0$ instead of assuming $\Theta=0$. As a consequence, we have $\dot{Y}=0$ ( but $\dot{C}\neq0$) from
Eq.(\ref{50b}), which means that $Y=Y(r)$ and without the loss of generality, we have
\begin{equation}\label{58b}
Y=1.
\end{equation}
The field equations for this case will become
\begin{align}
&
\begin{aligned}[t]
8\pi \Bigg[\rho^M
+\frac{\alpha \mathcal{G}^2}{2}-\frac{\beta T}{2}-
\frac{\hat{\hat{\psi}}_{00}}{X^2}\Bigg]=\frac{1}{X^2}
\Bigg(\frac{\dot{C}}{C}\Bigg)^2-\Bigg(\frac{C^\prime}{C}\Bigg)^2+\frac{1}{C^2}-2\frac{C^{\prime\prime}}{C},\label{59b}
\end{aligned}
\\[\jot]
 &
\begin{aligned}[t]
8\pi \Bigg[-(1+\beta)q^{M}+\frac{\hat{\hat{\psi}}_{01}}{X}\Bigg]=\frac{2}{X}
\Bigg(\frac{\dot{C^\prime}}{C}-\frac{\dot{C}}{C}
\frac{X^\prime}{X}\Bigg),\label{60b}
\end{aligned}
\\[\jot]
 &
\begin{aligned}[t]
&8\pi \Bigg[\rho^M \beta+(1+\beta)P^M_r
-\frac{\alpha \mathcal{G}^2}{2}+\frac{\beta T}{2}-
\hat{\hat{\psi}}_{11}\Bigg]=-\frac{1}{X^2}
\Bigg[2\frac{\ddot{C}}{C}-\frac{\dot{C}}{C}\Bigg(2\frac{\dot{X}}{X}\\ &-\frac{\dot{C}}{C}\Bigg)
\Bigg]+\frac{C^\prime}{C}\Bigg(\frac{C^\prime}{C}+2\frac{X^\prime}{X}\Bigg), \label{61b}
\end{aligned}
\\[\jot]
 &
\begin{aligned}[t]
&8\pi \Bigg[\rho^M \beta+(1+\beta)P^M_\bot
-\frac{\alpha \mathcal{G}^2}{2}+\frac{\beta T}{2}-
\frac{\hat{\hat{\psi}}_{22}}{C^2}\Bigg]=-\frac{1}{X^2}\Bigg(\frac{\ddot{C}}{C}-\frac{\dot{X}}{X}
\frac{\dot{C}}{C}\Bigg)\\ &+\frac{X^\prime}{X}
\frac{C^\prime}{C}+\frac{X^{\prime\prime}}{X}+\frac{C^{''}}{C}, \label{62b}
\end{aligned}
\end{align}
where double-hat indicates that the expressions of $\psi_{ab}$ are evaluated after using the MGB gravity and cavity model $Y=1$. The non-zero
components of the Bianchi identities, $T^{{-\lambda\mu}{(eff)}}_{;\mu}=0$, are
\begin{equation}\label{63b}
\frac{1}{X}\Bigg[\dot\rho^{eff}+\frac{2\dot{C}}{C}(P^{eff}_\bot+\rho^{eff})\Bigg]+2\frac{(XC)^\prime}{XC}q^{eff}+q^{\prime{eff}}=0,
\end{equation}
\begin{equation}\label{64b}
\begin{split}
\frac{1}{X}\Bigg(\dot{q}^{eff}+2q^{eff}\frac{\dot{C}}{C}\Bigg)+\frac{X^\prime}{X}(P^{eff}_r+\rho^{eff})+P^{\prime{eff}}_r+2(P^{eff}_r-P^{eff}_\bot)\frac{C^\prime}{C}=0.
\end{split}
\end{equation}
In the specific geodesic condition, we have $X'= 0\rightarrow X=1$.
At this point, the field equations (\ref{59b})-(\ref{62b}) will become
\begin{align}
&
\begin{aligned}[t]
8\pi \Bigg[\rho^M
+\frac{\alpha \mathcal{G}^2}{2}-\frac{\beta T}{2}-
\tilde{\psi}_{00}\Bigg]=\Bigg(\frac{\dot{C}}{C}\Bigg)^2-\Bigg(\frac{C^\prime}{C}\Bigg)^2+\frac{1}{C^2}-2\frac{C^{\prime\prime}}{C},\label{65b}
\end{aligned}
\\[\jot]
 &
\begin{aligned}[t]
8\pi \Bigg[-\Bigg(1+\beta\Bigg)q^{M}+\tilde{\psi}_{01}\Bigg]=2
\frac{\dot{C}^\prime}{C},\label{66b}
\end{aligned}
\\[\jot]
 &
\begin{aligned}[t]
&8\pi \Bigg[\rho^M \beta+(1+\beta)P^M_r
-\frac{\alpha \mathcal{G}^2}{2}+\frac{\beta T}{2}-
\tilde{\psi}_{11}\Bigg]=-\Bigg[2\frac{\ddot{C}}{C}+\Bigg(\frac{\dot{C}}{C}\Bigg)^2
\Bigg]\\ &+\Bigg(\frac{C^\prime}{C}\Bigg)^2-\frac{1}{C^2}, \label{67b}
\end{aligned}
\\[\jot]
 &
\begin{aligned}[t]
8\pi\Bigg[\rho^M \beta+(1+\beta)P^M_\bot
-\frac{\alpha \mathcal{G}^2}{2}+\frac{\beta T}{2}-
\frac{\tilde{\psi}_{22}}{C^2}\Bigg]=-\frac{\ddot{C}}{C}+\frac{C^{\prime\prime}}{C}, \label{68b}
\end{aligned}
\end{align}
where tilde indicates that the expressions of $\psi_{ab}$ are evaluated after using the MGB gravity model as well as cavity model $X=1$ and $Y=1$. It follows from Eq.(\ref{29b}) that
\begin{equation}\label{69b}
2\pi\rho^{eff}=\frac{m}{C^3}+2\pi( P^{eff}_r-2P^{eff}_\bot),
\end{equation}
and Eq.(\ref{38b}) implies
\begin{equation}\label{70b}
2\pi\rho^{eff}=2\pi(P^{eff}_r-4P^{eff}_\bot)-\varepsilon.
\end{equation}
Therefore, from Eqs.(\ref{69b}) and (\ref{70b}),
a conformally flat spacetime ($\varepsilon=0$) having geodesic fluid particles with isotropic pressures
$P^{eff}_{r}=P^{eff}_\bot=P^{eff}$ provides
\begin{equation}\label{71b}
4\pi P^{eff}+\frac{m}{C^3}=0.
\end{equation}
The above equation implies that such models satisfy the Darmois conditions if they show absorbing dissipative energy behaviour, that is $q^{eff}<0$, otherwise we get $M=0$ and $m<0$. Moreover, the condition $q^{eff}<0$ shows two different possibilities, which are $q^m<-q^D$ or $q^D<-q^M$.
Thus, this model satisfy Darmois conditions according to the
behaviour of dissipation due to gravitational impacts of DM along with dissipation from baryonic matter.

\section{Models Obeying Darmois Conditions}
Here, we assume various fundamental analytical models, not dependent upon thin shells on either $\Sigma^{(e)}$ or $\Sigma^{(i)}$, but are compatible with the Darmois conditions. We discovered that the non-dissipative models are the simplest. So, assuming $q^{eff}=0$ with Eq.(\ref{60b}) and after
integration we obtain
\begin{equation}\label{72b}
X=\frac{\dot{C}}{j_1},
\end{equation}
here $j_1(t)$ is an integration function of $t$. Without loss of generality, we reparametrize $t$ and assume that
\begin{equation}\label{73b}
 \dot{C}_{{\Sigma}^{(i)}}=j_1,
\end{equation}
with
\begin{equation}\label{74b}
X_{\Sigma^{(i)}}=1.
\end{equation}
Equations (\ref{31b}) and (\ref{72b}) provide
\begin{equation}\label{75b}
\dot{C}_{\Sigma^{(i)}}=j_1=U.
\end{equation}
It can be noticed from the above equation that for all these models the velocity $U$ is same for all the particles present in the cluster fluid.
This point was already mentioned in the previous section in Eq.(\ref{55b}).
By substituting Eq.(\ref{72b}) in Eqs.(\ref{59b}), (\ref{61b}) and (\ref{62b}) we get, by using
$D_T=\frac{1}{X} \frac{\partial}{\partial t}$ with Eqs.(\ref{72b}) and (\ref{73b})
\begin{align}
&
\begin{aligned}[t]
8\pi\Bigg[\rho^M
+\frac{\alpha \mathcal{G}^2}{2}-\frac{\beta T}{2}-
\frac{\hat{\hat{\psi}}_{00}}{X^2}\Bigg]=-\frac{1}{C^2}
\Big(2CC^{{\prime\prime}^2}+{C^\prime}^2-{\dot{C_{\Sigma^{(i)}}^2}}
-1\Big),\label{76b}
\end{aligned}
\\[\jot]
 &
\begin{aligned}[t]
&8\pi \Bigg[\rho^M \beta+(1+\beta)P^M_r
-\frac{\alpha \mathcal{G}^2}{2}+\frac{\beta T}{2}-
\hat{\hat{\psi}}_{11}\Bigg]=\frac{1}{C^2\dot{C}}\\ & D_T
\Bigg[\frac{1}{C^2\dot{C}}  D_T\Bigg(C(C^\prime)^2-{\dot{C_{\Sigma^{(i)}}^2}
-1}\Bigg)\Bigg],\label{77b}
\end{aligned}
\\[\jot]
 &
\begin{aligned}[t]
&8\pi \Bigg[\rho^M \beta+(1+\beta)P^M_\bot
-\frac{\alpha \mathcal{G}^2}{2}+\frac{\beta T}{2}-
\frac{\hat{\hat{\psi}}_{22}}{C^2}\Bigg]=\frac{1}{2C\dot{R}_{\Sigma^{(i)}}}
\\ &D_T \Big(2CC^{\prime
\prime}+ C^{\prime^2}-\dot{C^2}_{\Sigma^{(i)}}-1\Big). \label{78b}
\end{aligned}
\end{align}
It can be observed from Eqs.(\ref{76b}) and (\ref{78b}) that
\begin{equation}\label{79b}
{P^{eff}_\bot}=-\frac{D_T(\rho^{eff}
C^2)}{2C\dot{C}_{\Sigma^{(i)}}}.
\end{equation}
Calculating the mass function (\ref{29b}) by using Eqs.(\ref{58b}) and (\ref{72b}), it provides
\begin{equation}\label{80b}
m=-\frac{C}{2}({{C^\prime}^2}-\dot{C^2}_{\Sigma^{(i)}}-1),
\end{equation}
which turns Eq.(\ref{77b}) into
\begin{equation}\label{81b}
4\pi P^{eff}_{r}=-\frac{\dot{m}}{C^2 \dot{C}}.
\end{equation}
This model satisfies the junction conditions $P^{eff}_{r} {^{{\Sigma^{(e)}}}_{=}}0$ and
$P^{eff}_{r} {^{{\Sigma^{(i)}}}_{=}}0$ with $m{^{{\Sigma^{(i)}}}_{=}}0$ and $m{^{{\Sigma^{(e)}}}_{=}}M=constant$. By using Eq.(\ref{72b}) in Eq.(\ref{37b}), we get
\begin{equation}\label{82b}
\varepsilon=\frac{C}{4\dot{C}_{\Sigma^{(i)}}} D_T
\left(\frac{1}{C^2}
(2CC^{\prime\prime}-{{C^\prime}^2}+{\dot{C^2}_{\Sigma^{(i)}}})+1
\right).
\end{equation}
We will now explore some specific cases.

\subsection{Flat Conformal Models}
If we suppose that the geometry between $r=r_{\Sigma^{(e)}}$
and $r=r_{\Sigma^{(i)}}$ is conformally flat
$\varepsilon=0$, then Eq.(\ref{82b}) implies
\begin{equation}\label{83b}
2CC^{\prime\prime}+{{C^\prime}^2}-d_1 C^2+\dot{C^2}_{\Sigma^{(i)}} +
1=0,
\end{equation}
here $d_1(r)$ is an integrating function of $r$. The integration of above equation implies
\begin{equation}\label{84b}
{{C^\prime}^2}=C\left(\int d_1
dC+j_2\right)+\dot{C^2_{\Sigma^{(i)}}}+1,
\end{equation}
where $j_2(t)$ is an integrating function of $t$. By comparing
Eq.(\ref{80b}) and Eq.(\ref{84b}), we have
\begin{equation}\label{85b}
m=-\frac{C^2}{2}\left(\int d_1 dC+j_2\right),
\end{equation}
thus, $j_2(t)$ can be calculated through the junction condition
(\ref{46b}), providing
\begin{equation}\label{86b}
j_2{^{{\Sigma^{(i)}}}_{=}}-\int d_1 dC.
\end{equation}
Therefore, all these models are described through a single function
$d_1(r)$ and the selection of this function entirely depends on whether the remaining Darmois
conditions are fulfilled or not. Moreover, the evolution of all isotropic
fluids and spherically symmetric conformally flat spacetimes (in the
absence of dissipation) is shear free, but this is inaccurate
for anisotropic fluids \cite{50}. Hence, the models
investigated here are certainly considered anisotropic.

\subsection{Tangential Stress-less Models}
Let us assume $P^{eff}_\bot=0$, then after integrating Eq.(\ref{79b}), we obtain
\begin{equation}\label{87b}
\rho^{eff}=\frac{d_2}{C^2},
\end{equation}
where $d_2(r)$ is a function of $r$. Using
Eq.(\ref{87b}) into Eq.(\ref{76b}) yields
\begin{equation}\label{88b}
2CC^{\prime\prime}+{{C^\prime}^2}+8\pi
d_2-\dot{C^2}_{\Sigma^{(i)}}-1=0,
\end{equation}
and then with Eq.(\ref{80b}) it becomes
\begin{equation}\label{89b}
m^\prime=4\pi d_2 C^\prime.
\end{equation}
In order to get the models, a particular type of energy density or mass
function must be considered. We take the following condition into account
\begin{equation}\label{90b}
d_2=b_1=constant>0,
\end{equation}
Equations (\ref{34b}) and (\ref{87b}) imply
\begin{equation}\label{91b}
m=4\pi b_1 (C-C_{\Sigma^{(i)}}),
\end{equation}
and
\begin{equation}\label{92b}
M=4\pi b_1\Big(C_{\Sigma^{(e)}}-C_{\Sigma^{(i)}}\Big),
\end{equation}
\begin{equation}\label{93b}
\dot{C}_{\Sigma^{(i)}}=\dot{C}_{\Sigma^{(e)}},~~ X_{\Sigma^{(e)}}=X_{\Sigma^{(i)}}=1.
\end{equation}
It follows from Eqs.(\ref{81b}),(\ref{87b}) and (\ref{91b}) that
\begin{equation}\label{94b}
P^{eff}_r=8\pi\rho^{eff}
\Big(\frac{\dot{C}_{\Sigma^{(e)}}}{\dot{C}}-1\Big).
\end{equation}
Next, using Eq.(\ref{92b}) into Eq.(\ref{80b}) we get
\begin{equation}\label{95b}
C{{C^\prime}^2}=\alpha_1 C+\alpha_2,
\end{equation}
where
\begin{equation}\label{96b}
\alpha_1(t)=\dot{C^2}_{{\Sigma^{(i)}}}+1-8\pi b_1,~~ \alpha_2(t)=8\pi
b_1{C_{\Sigma^{(i)}}},
\end{equation}
and after integration
\begin{equation}\label{97b}
\Big[\alpha_1 C\Big(\alpha_1
C+\alpha_2\Big)\Big]^\frac{1}{2}-\alpha_2
ln\Big[\Big(\alpha_1R\Big)^\frac{1}{2}+\Big(\alpha_1
C+\alpha_2\Big)^\frac{1}{2}\Big]=\alpha_1^\frac{2}{3}\Big[r-r_0
\Big],
\end{equation}
where $r_0(t)$ is an integration function of $t$. Evaluating Eq.(\ref{97b}) on $\Sigma^{(i)}$ we get
\begin{equation}\label{98b}
\begin{split}
&\Big[{\Big(\alpha_1({\dot{C}_{\Sigma^{(i)}}}+1)\Big)^\frac{1}{2}-8\pi
b_1ln\Big(\alpha^\frac{1}{2}_1+\Big({\dot{C}^2_{\Sigma^{(i)}}+1}\Big)^\frac{1}{2}\Big)-4\pi
b_1ln C}\Big]C\\&
{^{{\Sigma^{(i)}}}_{=}}\alpha^\frac{3}{2}_1\Big(r-r_0\Big).
\end{split}
\end{equation}
This is the first order non-linear equation for $C_{\Sigma^{(i)}}$ which can be determined for any function $r_0(t)$. The solution of this integration, combined with Eq.(\ref{97b}) provides the complete information required to find $t$ and $r$ for all physical and
metric variables. It can be noticed that the energy density in the fluid corrections remains positive as well as regular. Furthermore,
the condition $r_0(t)$ with $0<\frac{\dot{C}_{\Sigma^{(i)}}}{C}-1\leq1$ assured the presence of positive
pressure which is less than the energy density.

\section{Graphical Analysis}
Now we will discuss the physical significance of $f(\mathcal{G}, T)$ gravity model on stellar bodies. For this reason, we take a static
spherical spacetime into consideration and it is given as
\begin{equation}\label{1a}
ds^2 = -X^2(r)dt^2 + Y^2(r)dr^2 + C^2(r)d\theta^2 + C^2(r) \sin^2 \theta d\phi^2.
\end{equation}
In order to discuss the physical behaviour of stellar bodies, we assume that the cluster of stars is formed of compact objects, for example, neutron stars and white dwarfs. We consider a stellar distribution $4U1820-30$ in an attempt to investigate the behaviour of $f(\mathcal{G}, T)$ gravity model. In order
to model a neutron star, a simple class model is considered depending upon spatial astral density (an idea which is analogous to de Vaucoulour's account in the exterior region, but not in the center). The idea of these class models was first given by Jaffe \cite{jaff} and Hernquist \cite{hern}, which have central astral densities equivalent to $r^{-2}$ and $r^{-1}$. Furthermore, these models can be implored to density profiles with a variety of central slopes given by
\begin{equation}\nonumber
\rho(r)=\frac{(3-\gamma)M\tilde{\alpha}}{4\pi r^\lambda (\tilde{\alpha}+r)^{4-\lambda}},
\end{equation}
where $\tilde{\alpha}$ expresses the scaling radius
and $M$ represents the whole mass which is
proportional to $r^{-\gamma}$ in the center.
The values of $\gamma$ are bounded by the interval
$[0, 3)$. Moreover, $\gamma=1$ represents Hernquist model
and $\gamma=2$ shows Jaffe model.
Nonetheless, we will use the Hernquist model for $\gamma=1$
in this study. We choose the metric functions in the form of
Krori-Barua ansatz \cite{kror1} as
\begin{equation}\label{1c}
X=e^a, \; Y=e^b,\; C=r,
\end{equation}
where $a=\tilde{B}r^2+\tilde{C}$ and $b=\tilde{A}r^2$. To model a celestial body, the Schwarzschild metric, as
exterior geometry, is employed to represent a realistic compact star model with a static
and asymptotically flat region, given as
\begin{equation}\label{1d}
ds^{2}=\left(1-\frac{2 M}{r}\right) d t^{2}-\left(1-\frac{2 M}{r}\right)^{-1} d r^{2}-r^{2}\left(d \theta^{2}+\sin ^{2} \theta d \phi^{2}\right).
\end{equation}
According to junction condition at $r=R$, we get
\begin{equation}\label{1e}
g_{t t}^{-}=g_{t t}^{+}, \quad g_{r r}^{-}=g_{r r}^{+}, \quad \frac{\partial g_{t t}^{-}}{\partial r}=\frac{\partial g_{t t}^{+}}{\partial r},
\end{equation}
where the $+$ and $-$ signs represent the exterior and interior surface of the compact star. By using Eq.(\ref{1e}) we get
\begin{eqnarray}
  \tilde{A} &=& -\frac{1}{R^2}\ln\bigg(1-\frac{2M}{R} \bigg), \label{65h} \\
  \tilde{B} &=& \frac{M}{R^3}\bigg(1-\frac{2M}{R} \bigg)^{-1}, \label{65i} \\
  \tilde{C} &=& \ln\bigg(1-\frac{2M}{R}\bigg)-\frac{M}{R}\bigg(1-\frac{2M}{R} \bigg)^{-1}. \label{65j}
\end{eqnarray}
The relation of mass-radius of compact stars has been used by many researchers to analyse the interior metric functions ($A$, $B$ and $C$) in accordance with the observational data \cite{bhar}. Rossi X-ray timing explorer accumulated the data based on satellite observations with reference to the composition of $4U1820-30$. The mass of this stellar body is found to be $2.25M_\odot$ which contains a high concentration of DM. Zhang \textit{et al.} \cite{zhan} investigated a neutron star from globular cluster binary system $4U1820-30$ and discovered mass of the direction as $2.25M_\odot$. Furthermore, Guver \textit{et al.} \cite{guve} also explored $4U1820-30$ with contrasting upper bound limits and observed $M=1.58\pm 0.06M_\odot$ and $R=9.11\pm 0.4 \ km$ with $1\sigma$ error. The values of $\tilde{A}$, $\tilde{B}$ and $\tilde{C}$ for mass $M=2.25 \ M_\odot$ and radius $R=10 \ km$ are given as
\begin{eqnarray}
  \tilde{A} &=& 0.010906441192 \ km^{-2}, \\
  \tilde{B} &=& 0.0098809523811 \ km^{-2},\\
  \tilde{C} &=& -2.0787393571141 \ km^{-2}.
\end{eqnarray}
Using the above values in field equations (\ref{25b})-(\ref{28b}),
the behaviors of seen matter dynamical variables (like density, radial pressure and tangential pressure)
are plotted according to parameter $\alpha$ and radius of stellar. Here, we will be considering $\beta=1$ for graphical representation.
\begin{figure}
\centering
\includegraphics[width=8cm]{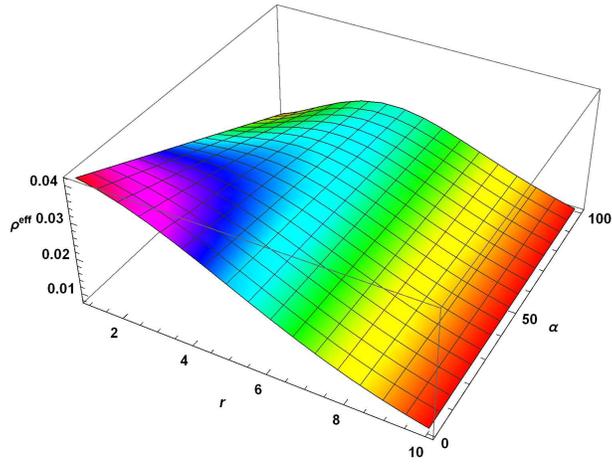}
\caption{Behaviour of $\rho^{eff}$.}
 \label{D}
\end{figure}
\begin{figure}
\centering
\includegraphics[width=8cm]{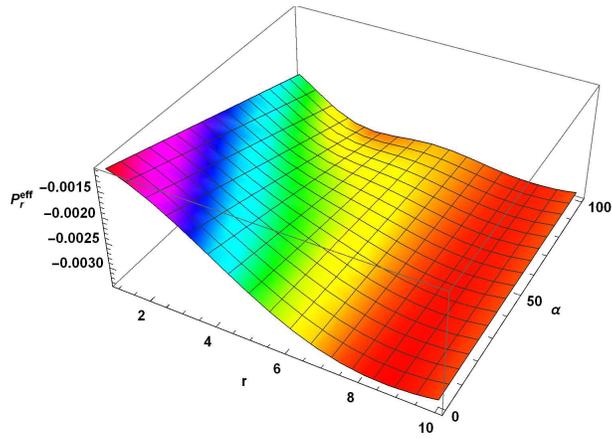}
\caption{Behaviour of $P_r^{eff}$.}
 \label{PR}
\end{figure}
\begin{figure}
\centering
\includegraphics[width=8cm]{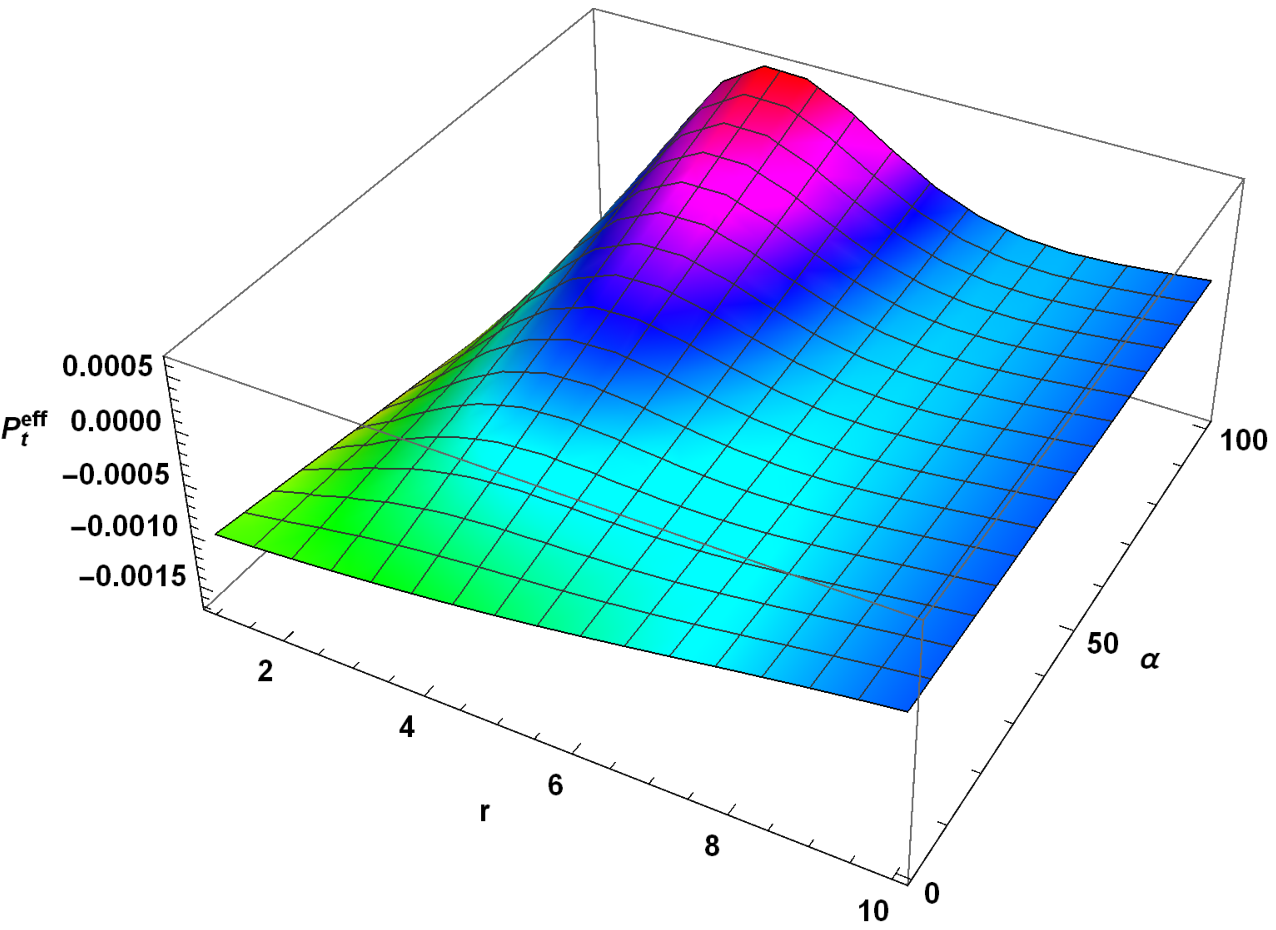}
\caption{Behaviour of $P_\bot^{eff}$.}
 \label{PT}
\end{figure}
It can be seen from Figures (\ref{D}), (\ref{PR}) and (\ref{PT}) that the behavior
of density and radial pressure is increasing towards the center, whereas
the behavior of tangential pressure is increasing towards the surface.
It is to mention here that, the parameter $\alpha$ indicates the DM role in the analysis
of cluster evolution. In this context, it can be observed from the graphs that, at the center,
the density is decreasing as the value of $\alpha$ is increasing. The radial pressure
shows the same behavior as density while tangential pressure depicts an
increasing behaviour with increasing values of $\alpha$.

Now we will describe the physical interpretation of the solutions obtained for cavity models. For this aim, we combine our results with the data obtained for stellar distribution $4U1820-30$.
\subsection{Case 1}
We consider $X=e^{\tilde{B}r^2+\tilde{C}}$, $Y=1$ and $C=r$ with $\tilde{B}=0.0098809523811 \ km^{-2}$ and $ \tilde{C}=-2.0787393571141 \ km^{-2}$. The graphs of Eqs.(\ref{59b})-(\ref{62b}) are plotted as functions of $r$ and $\beta$ since the parameter $\alpha$ vanishes for cavity model $Y=1$.
\begin{figure}
\centering
\includegraphics[width=8cm]{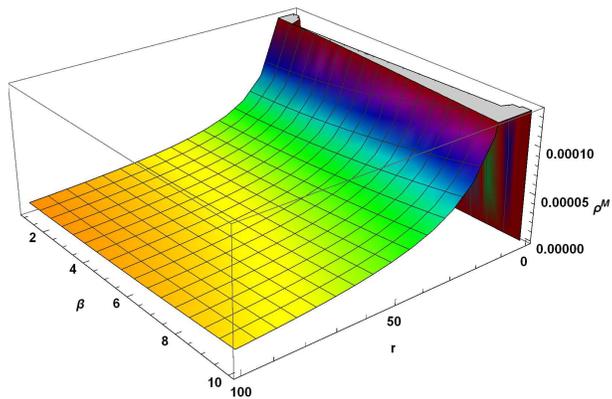}
\caption{Behaviour of $\rho^{M}$.}
 \label{CD}
\end{figure}
\begin{figure}
\centering
\includegraphics[width=8cm]{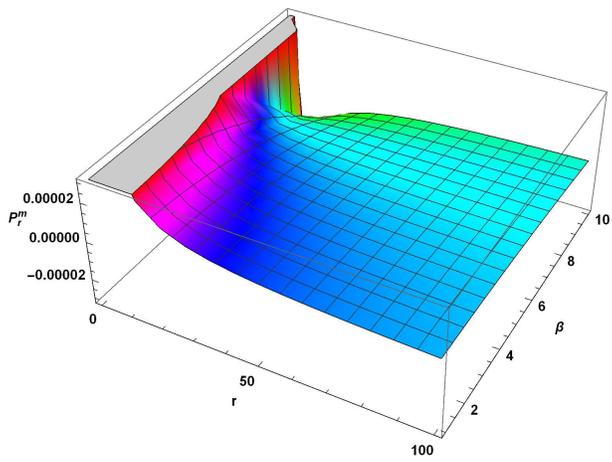}
\caption{Behaviour of $P_r^{M}$.}
 \label{CR}
\end{figure}
\begin{figure}
\centering
\includegraphics[width=8cm]{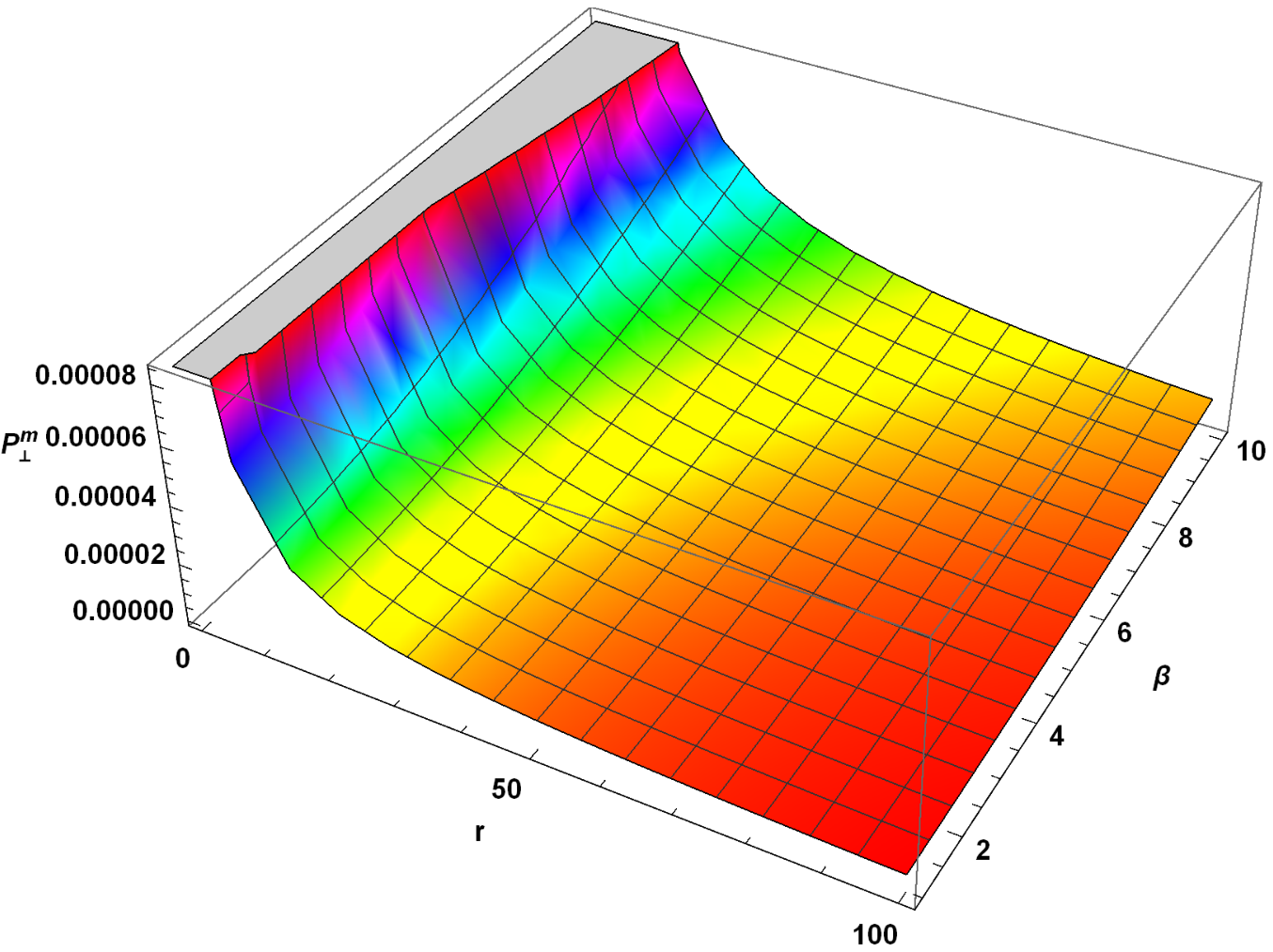}
\caption{Behaviour of $P_\bot^{M}$.}
 \label{CT}
\end{figure}
Figures (\ref{CD}), (\ref{CR}) and (\ref{CT}) indicate the behaviour of matter density and pressure profiles for cavity model $Y=1$. It can be seen from figure (\ref{CD}) that matter density vanishes as it moves towards the center, indicating there is no density present at the center in the cavity. Figures (\ref{CR}) and (\ref{CT}) show that the radial and tangential pressure are negligible at the center. Therefore, cavity model $Y=1$ shows no sign of density and pressure within the center of the cavity.
\subsection{Case 2}
Now we will discuss the behaviour of density and pressures for cavity model $X=1$ and $Y=1$. For this purpose, we consider the stellar distribution $4U1820-30$ with $X=1$, $Y=1$ and $C=r$. By using these conditions, density, radial and tangential pressure vanish at every point ($\rho^{M}, P_r^{M}, P_\bot^{M}=0$). This indicates that there is no density and pressure throughout the cavity, making it a vacuum cavity model.

\section{Summary and Discussion}
In this paper, we have studied the impact the MGB gravity on the dynamics of evolving cavity in cluster of stars. The evolving stars cluster are significant for the study of voids as well as galactic filaments. Voids are the regions with low density in a large scale matter distribution in the universe \cite{99}- \cite{104}. Moreover, voids of different scales have been found in our universe \cite{tikh, rud}. It should also be considered that voids are neither spherical nor empty, either in deep redshift surveys or in simulations. Nevertheless, for the sake of simplicity they are usually described as vacuum spherical cavities neighboured by a fluid. As a matter of fact, voids are the successor of the cavity model described by purely areal evolution condition. We have studied the outcomes appearing from the purely evolution condition. Moreover, it has been proved that this type of condition is specifically appropriate for the study of cavity evolution among the cluster of stars.

In order to study the evolution of cavity, we have considered a self-gravitating spherically symmetry geometry in context of MGB gravity, whereas, this higher order gravity theory is used as a tool to analyze the influence of DM on the cavity in cluster of stars. We have applied the purely evolution condition on the cavity surrounding the center. All dynamical equations have been obtained by using this condition. Furthermore, we also have discussed some cavity models by applying various conditions like expansion-free, conformally flat and geodesic. These models satisfy Darmois junction conditions in the presence of matter and DM on both hypersurfaces $\sum^{(i)}$ and $\sum^{(e)}$. We have also analysed the physical behaviour of cavity models by relating them with the date of stellar body $4U1820-30$ and investigated the behaviour of density as well as pressure profiles within the cavity. We concluded that the DM plays a vital role in controlling the evolution of cavity in cluster of stars.

\section*{Acknowledgments}
R. Manzoor, M. A. Sadiq  are thankful to the Higher Education Commission, Islamabad, Pakistan for its financial support under the grant No: Ref No. 20-15502/NRPU/R\& D/HEC/2021 2021.

\section*{Appendix}
\begin{equation}\nonumber
\begin{split}
\psi_{00}&=-\frac{4X^2}{C^2Y^2}f_{\mathcal{G},11}-\frac{8\dot{X}C'^2}{Y^2C^2X}f_{\mathcal{G},0}+\frac{8\dot{X}\dot{C}^2}{X^3C^2}f_{\mathcal{G},0}+\frac{12\dot{C}^2\dot{Y}}{X^2HC^2}f_{\mathcal{G},0}-\frac{16Y'^2X^2C'}{Y^6C}f_{\mathcal{G},1} \\ & -\frac{4C'^2X^2Y'}{Y^5C^2}f_{\mathcal{G},1}+\frac{16\dot{X}Y'C'}{CAH^3}f_{\mathcal{G},0}+\frac{8\dot{C}Y'C'}{C^2Y^3}f_{\mathcal{G},0}-\frac{16Y'\dot{C}\dot{Y}}{Y^4C}f_{\mathcal{G},1}-\frac{8C'\dot{C}\dot{Y}}{C^2Y^3}f_{\mathcal{G},1} \\ & +\frac{16\dot{X}\dot{C}\dot{Y}}{X^3CH}f_{\mathcal{G},0}-\frac{4\dot{C}^2}{C^2Y^2}f_{\mathcal{G},11}+\frac{4X^2C'^2}{C^2Y^4}f_{\mathcal{G},11}+\frac{8\dot{X}}{C^2X}f_{\mathcal{G},0}-\frac{4\dot{Y}C'^2}{C^2Y^3}f_{\mathcal{G},0} \\ & -\frac{4Y'\dot{C}^2}{C^2Y^3}f_{\mathcal{G},1} -\frac{4Y'X^2}{C^2Y^3}f_{\mathcal{G},1}+\frac{4\dot{Y}}{HC^2}f_{\mathcal{G},0}-\frac{8\dot{C}C''}{Y^2C^2}f_{\mathcal{G},0}+\frac{16Y'X^2C''}{Y^5C}f_{\mathcal{G},1} \\ & -\frac{16\dot{X}C''}{CH^2X}f_{\mathcal{G},0} +\frac{8C'X^2C''}{Y^4C^2}f_{\mathcal{G},1}
\end{split}
\end{equation}

\begin{equation}\nonumber
\begin{split}
\psi_{01}&=-\frac{8\dot{C}'\dot{C}}{C^2X^2}f_{\mathcal{G},0}+\frac{8\dot{C}'C'}{C^2Y^2}f_{\mathcal{G},1}+\frac{16\dot{C}'Y'}{Y^3C}f_{\mathcal{G},1}
-\frac{16\dot{C}'\dot{X}}{X^3C}f_{\mathcal{G},0}-\frac{4C'^2X'}{C^2AH^2}f_{\mathcal{G},0} \\
& -\frac{12C'^2\dot{Y}}{C^2Y^3}f_{\mathcal{G},1}+\frac{12\dot{C}^2X'}{C^2X^3}f_{\mathcal{G},0}+\frac{4\dot{C}^2\dot{Y}}{C^2X^2Y}f_{\mathcal{G},1}-\frac{4}{C^2}f_{\mathcal{G},01}-\frac{4\dot{C}^2}{C^2X^2}f_{\mathcal{G},01} \\ & +\frac{4X'}{AC^2}f_{\mathcal{G},0}+\frac{4\dot{Y}}{C^2Y}f_{\mathcal{G},1}+\frac{4C'^2}{C^2Y^2}f_{\mathcal{G},01}+\frac{16\dot{X}\dot{C}X'}{CA^4}f_{\mathcal{G},0}-\frac{16Y'C'\dot{Y}}{CH^4}f_{\mathcal{G},1} \\ & -\frac{16\dot{C}X'Y'}{Y^3AC}f_{\mathcal{G},1}+\frac{16\dot{Y}C'\dot{X}}{X^3CH}f_{\mathcal{G},0}-\frac{8\dot{C}X'C'}{C^2AH}f_{\mathcal{G},1}+\frac{8\dot{Y}C'\dot{C}}{C^2X^2Y}f_{\mathcal{G},0}
\end{split}
\end{equation}

\begin{equation}\nonumber
\begin{split}
\psi_{11}&=-\frac{8C'^2Y'}{Y^3C^2}f_{\mathcal{G},1}-\frac{12C'^2X'}{Y^2AC^2}f_{\mathcal{G},0}+\frac{8Y'\dot{C}^2}{X^2C^2Y}f_{\mathcal{G},1}+\frac{\dot{X}Y^2\dot{C}^2}{X^5C^2}f_{\mathcal{G},0}+\frac{16\dot{X}^2Y^2\dot{C}}{X^6C}f_{\mathcal{G},0} \\ & +\frac{16Y'\ddot{C}}{CA^2Y}f_{\mathcal{G},1}-\frac{16\dot{X}Y^2\ddot{C}}{X^5C}f_{\mathcal{G},0}-\frac{8\dot{C}Y^2\ddot{C}}{X^4C^2}f_{\mathcal{G},0}-\frac{4Y^2}{C^2X^2}f_{\mathcal{G},00}+\frac{8C'\ddot{C}}{X^2C^2}f_{\mathcal{G},1} \\ & -\frac{4Y^2\dot{C}^2}{C^2X^4}f_{\mathcal{G},00}+\frac{4C'^2}{C^2X^2}f_{\mathcal{G},00}+\frac{8Y'}{C^2Y}f_{\mathcal{G},11}+\frac{4\dot{X}C'^2}{C^2X^3}f_{\mathcal{G},0}+\frac{4X'\dot{C}^2}{C^2X^3}f_{\mathcal{G},1} \\ & +\frac{4X'}{AC^2}f_{\mathcal{G},1}-\frac{4\dot{X}Y^2}{C^2X^3}f_{\mathcal{G},0}-\frac{16Y'\dot{X}\dot{C}}{CA^3Y}f_{\mathcal{G},1}-\frac{8C'\dot{X}\dot{C}}{C^2X^3}f_{\mathcal{G},1}-\frac{16X'Y'C'}{CH^3X}f_{\mathcal{G},1} \\ & +\frac{16\dot{X}X'C'}{X^4C}f_{\mathcal{G},0}+\frac{8\dot{C}X'C'}{C^2X^3}f_{\mathcal{G},0}
\end{split}
\end{equation}

\begin{equation}\nonumber
\begin{split}
\psi_{22}&=-\frac{8\dot{Y}^2C\dot{C}}{Y^2X^4}f_{\mathcal{G},0}+\frac{4C^2X'Y'}{AH^5}f_{\mathcal{G},11}-\frac{4C^2\dot{X}\dot{Y}}{X^3Y^3}f_{\mathcal{G},11}-\frac{4C\dot{X}\dot{C}}{X^3Y^2}f_{\mathcal{G},11}-\frac{4CH'C'}{X^2Y^3}f_{\mathcal{G},00} \\ & -\frac{4CA'C'}{AH^4}f_{\mathcal{G},11}-\frac{4C\dot{C}Y}{X^4Y}f_{\mathcal{G},00}+\frac{8C\dot{C}\dot{Y}}{X^2Y^3}f_{\mathcal{G},11}+\frac{8CH'C'}{Y^5}f_{\mathcal{G},11}+\frac{8\dot{X}^2C^2\dot{Y}}{X^6Y}f_{\mathcal{G},0} \\ & +\frac{8\dot{X}^2C\dot{C}}{X^6}f_{\mathcal{G},0}-\frac{8C\dot{C}X'^2}{X^4Y^2}f_{\mathcal{G},0}+\frac{8\dot{C}CA'}{X^3Y}f_{\mathcal{G},01}-\frac{8C\dot{C}^2C'}{X^2Y^4}f_{\mathcal{G},1}+\frac{8C\dot{Y}C'}{X^2Y^3}f_{\mathcal{G},01} \\ & +\frac{4CC'\ddot{Y}}{X^2Y^3}f_{\mathcal{G},1}-\frac{8\dot{X}C^2\ddot{Y}}{X^5Y}f_{\mathcal{G},0}-\frac{4\dot{C}C\ddot{Y}}{X^4Y}f_{\mathcal{G},0}-\frac{4C'CA''}{Y^4X}f_{\mathcal{G},1}+\frac{8\dot{X}C^2X''}{X^4Y^2}f_{\mathcal{G},0} \\ & +\frac{8\dot{Y}CC''}{X^2Y^3}f_{\mathcal{G},0}+\frac{4\dot{C}CA''}{Y^2X^3}f_{\mathcal{G},0}+\frac{8C\dot{C}'\dot{Y}}{X^2Y^3}f_{\mathcal{G},1}+\frac{8C\dot{C}'X'}{X^3Y^2}f_{\mathcal{G},0}-\frac{8CC''}{Y^4}f_{\mathcal{G},11} \\ & -\frac{8\dot{Y}CH'C'}{Y^4X^2}f_{\mathcal{G},0}+\frac{4C'CA'Y}{Y^5X}f_{\mathcal{G},1}+\frac{8\dot{X}CA'C'}{X^4Y^2}f_{\mathcal{G},0}-\frac{4\dot{Y}CA'C'}{X^3Y^3}f_{\mathcal{G},0}-\frac{4X'C\dot{C}\dot{Y}}{X^3Y^3}f_{\mathcal{G},1} \\ & -\frac{8\dot{X}C\ddot{C}}{X^5}f_{\mathcal{G},0}+\frac{4C\ddot{C}}{X^2Y^2}f_{\mathcal{G},11}+\frac{4C^2\ddot{Y}}{X^2Y^3}f_{\mathcal{G},11}+\frac{4CC''}{X^2Y^2}f_{\mathcal{G},00}-\frac{4C^2X''}{AH^4}f_{\mathcal{G},11} \\ & -\frac{8C\dot{C}'}{X^2Y^2}f_{\mathcal{G},01}-\frac{4C'C\dot{X}\dot{Y}}{X^3Y^3}f_{\mathcal{G},1}-\frac{8\dot{X}C^2X'Y'}{X^4Y^3}f_{\mathcal{G},0}-\frac{4\dot{C}CA'Y'}{X^3Y^3}f_{\mathcal{G},0}+\frac{4\dot{X}C\dot{C}\dot{Y}}{X^5Y}f_{\mathcal{G},0} \\ & +\frac{4Y'^2C^2X'}{Y^6X}f_{\mathcal{G},1}+\frac{4\dot{Y}^2C^2\dot{X}^2}{Y^2X^5}f_{\mathcal{G},0}-\frac{4\dot{Y}C\ddot{C}}{HA^4}f_{\mathcal{G},0}+\frac{4Y'C^2\ddot{Y}}{Y^4X^2}f_{\mathcal{G},1}-\frac{4\dot{Y}C^2\ddot{Y}}{Y^2X^4}f_{\mathcal{G},0} \\ & -\frac{4Y'C^2X''}{Y^5X}f_{\mathcal{G},1}-\frac{4X'CC''}{AH^4}f_{\mathcal{G},1}+\frac{4\dot{X}CC''}{Y^2X^3}f_{\mathcal{G},0}+\frac{4\dot{Y}C^2X''}{X^3Y^3}f_{\mathcal{G},0}+\frac{4Y'C\ddot{C}}{X^2Y^3}f_{\mathcal{G},1} \\ & -\frac{4\dot{Y}C^2X'Y'}{Y^4X^3}f_{\mathcal{G},0}-\frac{4Y'C\dot{X}\dot{C}}{X^3Y^3}f_{\mathcal{G},1}-\frac{4\dot{X}CH'C'}{X^3Y^3}f_{\mathcal{G},0}-\frac{4Y'C^2\dot{X}\dot{Y}}{Y^4X^3}f_{\mathcal{G},1}
\end{split}
\end{equation}


\begin{thebibliography}{40}
\bibitem{perl1} Perlmutter, S. \textit{et al.}: Astrophys. J. \textbf{517}(1999)565.
\bibitem{riess1} Riess, A.G. \textit{et al.}: Astrophys. J. \textbf{659}(2007)98.
\bibitem{koma1} Komatsu, E. \textit{et al.}: Astrophys. J. Suppl. \textbf{192}(2011)18.
\bibitem{1} Aghanim, N. \textit{et al.}: Astro. Astroph. \textbf{641}(2020)A6.
\bibitem{2} Nojiri, S. and Odintsov, S.D.: Int. J. Geom. Meth. Mod. Phys. \textbf{4}(2007)115.
\bibitem{3} Azadi, A., Momeni, D. and Nouri-Zonoz, M.: Phys. Lett. \textbf{B670}(2008)210.
\bibitem{4} Starobinsky, A.A.: J. Exp. Theor. Phy. Lett. \textbf{86}(2009)157.
\bibitem{5} Appleby, S., Battye, R. and Starobinsky, A. A.: JCAP \textbf{06}(2010)005.
\bibitem{6} Capozziello, S., De-Filippis, E. and Salzano, V.: Mon. Not. Roy. Astron. Soc. \textbf{394}(2009)947.
\bibitem{7} Bamba, K., Capozziella, S., Nojiri, S. and Odintsov, S.D.: Astrophys. Space Sci. \textbf{342}(2012)155.
\bibitem{8} Nojiri, S. and Odintsov, S.D.: \textit{Problems of Modern Theoretical Physics}, A Volume in honour of Prof. Buchbinder, I.L. in the occasion of his 60th birthday, p.266-285, (TSPU Publishing, Tomsk), arXiv:0807.0685.
\bibitem{9} Nojiri,S. and Odintsov, S.D.: Phy. Lett. \textbf{599}(2004)137.
\bibitem{10} Akbar, M. and Cai, R.: Phy. Lett. \textbf{B648}(2007)243.
\bibitem{11} Cai, R., Liu, Y. and Sun, Y.: JHEP \textbf{0910}(2009)080.
\bibitem{12} Harko, T., Lobo, F.S.N., Nojiri, S. and Odintsov, S.D.: Phys. Rev. D \textbf{84}(2011)024020.
\bibitem{13} Capozziello, S., Martin-Moruno, P. and Rubano, C.: Phys. Lett. B \textbf{664}(2008)12.
\bibitem{14} Felice, A.D. and Tsujikaswa, S.: Living Rev. Rel. \textbf{13}(2010)3.
\bibitem{15} Sharif, M. and Ikram, A.: Eur. Phys. J. C \textbf{76}(2016)640.
\bibitem{17} Cognola, G., Elizalde, E., Nojiri, S., Odintsov, S.D. and Zerbini, S.: Phys. Rev. D \textbf{73}(2006)084007.
\bibitem{18} Cognola, G., Elizalde, E., Nojiri, S., Odintsov, S.D, Sebastiani, L. and Zerbini, S.: Phys. Rev. D \textbf{77}(2008)046009.
\bibitem{19} Elizalde, E., Myrzakulov, R., Obukhov, V.V. and Saez-Gomez, D.: Class. Quant. Grav. \textbf{27}(2010)095007.
\bibitem{20} Cognola, G., Elizalde, E., Nojiri, S., Odintsov, S.D. and Zerbini, S.: Eur. Phys. J. C \textbf{64}(2009)483.
\bibitem{21} Nojiri, S. and Odintsov, S.D.: Phys. Lett. \textbf{B631}(2005)1.
\bibitem{22} Chiba, T.: J. Cosmol. Astropart. Phys. \textbf{03}(2005)008.
\bibitem{24} Shamir, M.F. and Ahmad, M.: Eur. Phys. J. C \textbf{77}(2017)55.
\bibitem{25} Sharif, M. and Ikram, A.: Phys. of Dark Uni. \textbf{17}(2017)1.
\bibitem{26} Sharif, M. and Ikram, A.: Ad. High Energy Phys. \textbf{2018}(2018)2563871.
\bibitem{28} Shamir, M.F.: Adv. High Energy Phys \textbf{2017}(2017)6378904.
\bibitem{sharhas} Shamir M.F. and Sadiq, M.A.: Eur. Phys. J. C \textbf{78}(2018)279; Commun. Theor. Phys.  \textbf{71}(2019)220.
\bibitem{27} Shamir, M.F. and Ahmad M.: Mod. Phys. Lett. A \textbf{32}(2017)1750086.
\bibitem{29} Bhatti, M.Z., Sharif, M. and Yousaf, Z.: Int. J. Mod. Phys. D \textbf{27}(2018)1850044.
\bibitem{30} Baumgardt, H. and Makino, J.: Mon. Not. Roy. Astron. Soc. \textbf{340}(2003)227.
\bibitem{isper} Ipser, J.R. and Thorne, K.S.: Astrophys. J. \textbf{154}(1968)251.
\bibitem{krui} Kruijssen, J.M.D. \textit{et al.}: Mon. Not. R. Astron. Soc. \textbf{414}(2011)1364.
\bibitem{krum} Krumholz, M.R., McKee, C.F. and Hawthorn, J.B.: Ann. Rev. of Astron. Astrophys. \textbf{57}(2019)227.
\bibitem{rubab} Manzoor, R. and Shahid, W.: Phys. Dar. Univ. \textbf{33}(2021)100844.
\bibitem{Wilt} Wiltshire, D.L.: \textit{Dark Matter in Astroparticle and Particle Physics} (2008)565-596.
\bibitem{hoyle} Hoyle, F. and Vogeley, M.S.: Astrophys. J. \textbf{607}(2004)751.
\bibitem{tikh} Tikhonov, A.V. and Karachentsev, I.D.: Astrophys. J. \textbf{653}(2006)969.
\bibitem{rud} Rudnick, L., Brown, S. and Williams, L.R.: Astrophys. J. \textbf{671}(2007)40.
\bibitem{31} Skripkin, Soviet, V.A.: Physics-Doklady \textbf{135}(1960)1183.
\bibitem{32} Herrera, L., Santos, N.O. and Wang, A.: Phys. Rev. D \textbf{78}(2008)084026.
\bibitem{33} Herrera, L., Le Denmat, G. and Santos, N.O.: Phys. Rev. D \textbf{79}(2009)087505.
\bibitem{34} Darmois, G.: Memorial des Sciences Mathematiques (Gauthier-Villars, Paris, 1927) Fasc. 25.
\bibitem{35} Herrera, L., Le Denmat, G. and Santos, N.O.: Class. Quantum. Grav. \textbf{27}(2010)135017.
\bibitem{36} Sharif, M. and Bhatti, M.Z.: Astrophys. Space Sci. \textbf{352}(2014)883.
\bibitem{37} Yousaf, Z. and Bhatti, M.Z.: Eur. Phys. J. C \textbf{76}(2016)1.
\bibitem{39} Yousaf, Z., Bamba, K. and Bhatti, M.Z.: Phys. Rev. D \textbf{93}(2016)064059.
\bibitem{cogn1} Cognola, G., Elizalde, E., Nojiri, S., Odintsov, S.D. and Zerbini, S.: Phys. Rev. D \textbf{75}(2007)086002.
\bibitem{42} Maartens, R. arXiv:astro-ph/9609119 (1996).
\bibitem{43} Herrera, L., Di Prisco, A. and Ospino, J.: Gen Relativ Gravit \textbf{42}(2010)1585.
\bibitem{44} Misner, C. and Sharp, D.: Phys. Rev. B \textbf{571}(1964)136.
\bibitem{45} Cahill, M. and McVittie, G.J.: Math. Phys. \textbf{11}(1970)1382.
\bibitem{41} Herrera, L. and Santos, N.O.: Phys. Rev. D \textbf{70}(2004)084004.
\bibitem{46} Santos N.O.: MNRAS \textbf{216}(1985)403.
\bibitem{47} Bonnor, B.W., Oliveira, A. and Santos, N.O.: Phys. Rep. \textbf{181}(1989)269.
\bibitem{48} Chan, R.: MNRAS \textbf{316}(2000)588.
\bibitem{49} Israel, W.: Nuovo Cimento B \textbf{44}(1966)1.
\bibitem{51} Kippenhahn, R. and Weigert, A.: \textit{Stellar Structure and Evolution} (Berlin: Springer) (1990).
\bibitem{50} Herrera L., Di Prisco A., Martin J., Ospino J., Santos, N.O. and Troconis, O.: Phys. Rev. D \textbf{69}(2004)084026.
\bibitem{jaff} Jaffe, W.: MNRAS. \textbf{202}(1983)995.
\bibitem{hern} Hernquist L.: ApJ. \textbf{356}(1990)359.
\bibitem{kror1} Krori, K. D. and Barua, J.: J. Phys. A. Math. Gen. \textbf{8}(1975)508.
\bibitem{bhar} Bhar, Piyali.: Eur. Phys. J C \textbf{79}(2019)138.
\bibitem{zhan} Zhang, W. \textit{et al.}: Astro Physical J. \textbf{L500}(1998)171.
\bibitem{guve} Guver, T. \textit{et al.}: Astro Physical J. \textbf{719}(2010)1807.
\bibitem{99} ZelDovich, Y.B.: Astron. and Astrophys. \textbf{500}(1970)13.
\bibitem{100} Peebles, P.J.E.: Astrophys. J., \textbf{557}(2001)495.
\bibitem{101} F. Hoyle and M.S. Vogeley: Astrophys. J. \textbf{566}(2002)641.
\bibitem{102} Odrzywolek, A.: Phys. Rev. D \textbf{80}(2009)103515.
\bibitem{103} Gaite, J.: JCAP \textbf{11}(2009)004.
\bibitem{104} Weygaert, R.V. and Platen, E.: Int. J. Mod. Phys. \textbf{01}(2011)41.


\end{thebibliography}
\end{document}